\documentclass[twoside,fleqn]{article}
\usepackage{graphicx}
\usepackage{espcrc2}

\def\spose#1{\hbox to 0pt{#1\hss}}
\def\ltapprox{\mathrel{\spose{\lower 3pt\hbox{$\mathchar"218$}}
 \raise 2.0pt\hbox{$\mathchar"13C$}}}
\def\gtapprox{\mathrel{\spose{\lower 3pt\hbox{$\mathchar"218$}}
 \raise 2.0pt\hbox{$\mathchar"13E$}}}
\def\inapprox{\mathrel{\spose{\lower 3pt\hbox{$\mathchar"218$}}
 \raise 2.0pt\hbox{$\mathchar"232$}}}

\newcommand{\etal}{{\em et al.}}

\newcommand{\NPB}{Nucl.Phys.B \ }

\newcommand{\PRD}{Phys.Rev.D \ }

\newcommand{\PRL}{Phys.Rev.Lett \ }

\newcommand{\be}{\begin{equation}}
\newcommand{\ee}{\end{equation}}
\newcommand{\bea}{\begin{eqnarray}}
\newcommand{\eea}{\end{eqnarray}}

\newcommand{\ksea}{\mbox{$\kappa_{\rm sea}$}}
\newcommand{\csw}{\mbox{$c_{\rm sw}$}}

\newcommand{\kval}{\mbox{$\kappa_{\rm val}$}}
\newcommand{\plus}{\makebox[5pt][c]{$+$}}
\newcommand{\minus}{\makebox[5pt][c]{$-$}}

\newcommand{\err}[2]{\raisebox{0.08em}{\scriptsize
                          {$\;\begin{array}{@{}l@{}}
                          \plus\makebox[1.55em][r]{#1} \\[-0.12em]
                          \minus\makebox[1.55em][r]{#2}
                        \end{array}$}}}

\title{UKQCD's latest results for the static quark potential and light hadron spectrum with $O(a)$ improved dynamical fermions.}
\author{UKQCD Collaboration, presented by J.~Garden
\address{Department of Physics \&\ Astronomy, University of Edinburgh,
The King's Buildings, EH9 3JZ (UK)}
\thanks{Supported by the PPARC grant GR/L22744}, 
}
\begin{document}
\begin{abstract}
We present UKQCD's latest results for the static quark potential and
light hadron spectrum obtained from matched simulations using two
flavours of dynamical quarks. We report that using matched ensembles
helps disentangle screening effects from discretisation errors in the
static quark potential.
\end{abstract}
\maketitle
\section{Introduction}
\noindent Previous simulations at fixed $\beta$ with several values of \ksea,
have shown a strong dependence on the lattice spacing as $\ksea$ is
varied~\cite{dynam}. This complicates the chiral extrapolations and
obscures comparisons with quenched simulations. UKQCD have proposed
that simulations should be carried out at fixed lattice spacing, $a$,
for different values of \ksea.  In this way it is possible to study
the effect of varying the sea quark mass at the same effective lattice
volume.

The lattice spacing is fixed by tuning the bare parameters, $\beta$
and \ksea. This is achieved by comparing a lattice observable with the
physical value. In our case the Sommer scale, $r_0$, ~\cite{r0} has been used,
where,
\be
 F(r_0/a)(r_0/a)^2 = 1.65, \qquad r_0 =
0.49 {\rm fm}
\ee
and $F(r_0/a)$ is the force between a static quark anti-quark
pair. The Sommer scale was selected as it can be determined with good
statistical precision and is independent of the valence quarks,
avoiding the need for extrapolations. The details of the matching
technique can be found in~\cite{tune}.

\section{Simulation parameters}
\noindent The simulations are performed with two flavours of dynamical fermions.
We use the standard Wilson gauge action together with the $O(a)$
improved Wilson fermion action. The clover coefficient used was
determined non-perturbatively by the {\sc Alpha} Collaboration.

All simulations were carried out on a $16^3 \times 32$ lattice. The
parameters for the matched ensembles are shown in
Table~\ref{parameters}. The last entry shows a simulation at the
lightest $\ksea$ which is not matched.
\begin{table}
\caption{Simulation parameters}
\begin{tabular*}{\columnwidth}{cccll} \hline \hline
$\beta$ &$\csw$ &$\ksea$ &$\kval$ &Conf. \\ \hline \hline
5.29    &1.92   &0.1340  &0.1335, 0.1340,  & 101\\
&       &                &0.1345, 0.1350   &    \\\hline
5.26    &1.95   &0.1345  &0.1335, 0.1340                   &101\\
&       &                &0.1345, 0.1350   &    \\\hline
5.2     &2.02   &0.1350  &0.1335, 0.1340,                  &150\\
&       &                &0.1345, 0.1350   &    \\ \hline       
5.9     &1.89   &Quen. &0.1325, 0.1330 & 100 \\
        &       &       &0.1335 & \\ \hline \hline 
\multicolumn{5}{l}{Lightest $\ksea$ simulation.\vspace{0.15cm}}\\ \hline \hline
5.2     &2.02   &0.1355  &0.1340, 0.1345,   &102\\ 
        &       &        &0.1350, 0.1355    &   \\ \hline \hline
\end{tabular*}
\label{parameters}
\vspace{-0.5cm}
\end{table}
Table~\ref{r0a} shows the results for the lattice spacing and
$r_0/a$ which were obtained using the method described
in~\cite{edwards}. This corresponds to an effective lattice volume of
approximately 1.7 fm for the matched simulations.

\begin{table}
\caption{Lattice spacing and $r_0/a$. The
statistical error is in parentheses and the second error is an
estimate of the systematic errors.}
\begin{center}
\begin{tabular*}{\columnwidth}{llll} \hline \hline
$\beta$ &$\ksea$ & $r_0/a$ &$a[\rm fm]$\\ \hline \hline
5.9   &Quen. &4.332(45)         &0.1131(12)\err{33}{120}\\
5.29    &.1340 &4.450(61)\err{29}{61}&0.1101(15)\err{13}{7}\\ 
5.26    &.1345 &4.581(59)\err{0}{120}&0.1070(14)\err{30}{0}\\
5.2     &.1350  &4.576(80)\err{14}{130}&0.1071(19)\err{40}{3}\\ \hline
5.2     &.1355  &4.914(82)\err{70}{19}          &0.0997(17)\err{40}{14}\\
\hline \hline
\end{tabular*}
\end{center}
\label{r0a}
\vspace{-0.5cm}
\end{table}
\section{Static quark potential} 
\begin{figure}
\begin{center}
    \leavevmode
    \includegraphics[angle=0,width=3in]{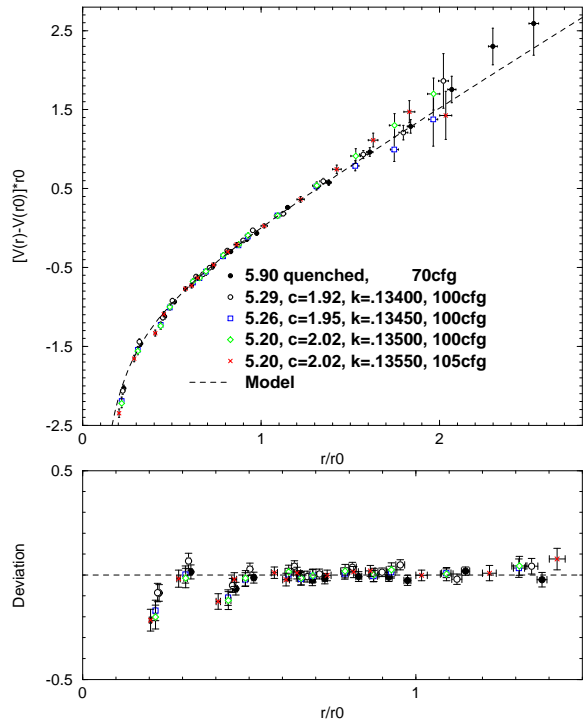}
  \end{center}
\vspace{-1cm}
\caption{Rescaled static potential on matched ensembles compared with the string model.}
\vspace{-0.5cm}
\label{pot}
\end{figure}

\noindent The standard form for the static quark potential 
\be
V(r) = V_0 + \sigma r - \frac{e}{r},
\ee 
can be rescaled in terms of $r_0$ as
\be
[V(r) - V(r_0)]r_0 = (1.65 - e)\left(\frac{r}{r_0} -
1\right) - e \left(\frac{r_0}{r} - 1\right)
\label{vr0}
\ee
Figure~\ref{pot} shows the results for the static quark potential
compared with eqn.~\ref{vr0}.  We observe good agreement with the
universal fit $\pi/12r + \sigma r$. With these results there is no
indication of string breaking at large $r/r_0$. However the plot of
the deviation from the model shows significant discretisation
errors. At short distances where the fits have to take this into
account, there is some evidence that the lighter quark data lie below
the heavier quark data.
\begin{figure}
\begin{center}
    \leavevmode
    \includegraphics[angle=270,width=3in]{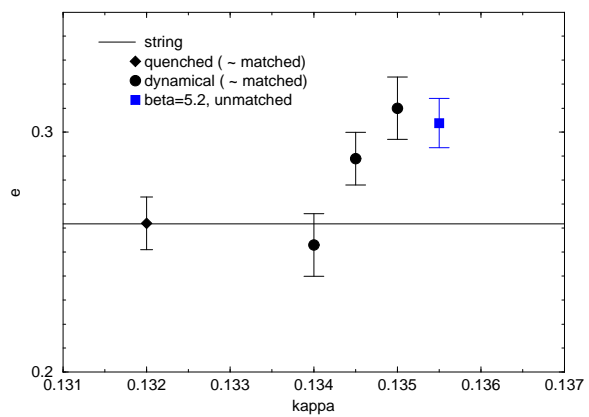}

  \end{center}
\vspace{-1cm}
\caption{Fitted values of the parameter $e$ as a function of $\ksea$. The solid line is the L\"{u}scher value $e = \frac{\pi}{12}$.}
\label{e}
\vspace{-0.5cm}
\end{figure}
Parametric fits for the $1/r$ coefficient $e$, see Figure~\ref{e}, show
an increase for the dynamical data of $15\% \pm 4\%$. This is
consistent with perturbation theory~\cite{khadra} which suggests an
increase for $e$ of around $14\%$ for ${\rm N_{\rm f}} = 2$.
\begin{table}
\caption{$m_{\rm PS}/m_{\rm V}$ ratios for dynamical data sets.}
\begin{center}
\begin{tabular}{ccc}\hline \hline
$\beta$ &$\ksea = \kval$ & $m_{\rm PS}/m_{\rm V}$ \\ \hline \hline
5.29    &0.1340  &0.830\err{8}{6}\\
5.26    &0.1345 &0.785\err{9}{9}\\
5.2     &0.1350 &0.693\err{11}{13} \\
5.2     &0.1355 & 0.584\err{25}{16}\\
\hline \hline
\end{tabular}
\end{center}
\label{ratio}
\vspace{-0.75cm}
\end{table}
\section{Light hadron spectrum} 
\begin{figure}
\begin{center}
    \leavevmode
    \includegraphics[angle=270,width=3in]{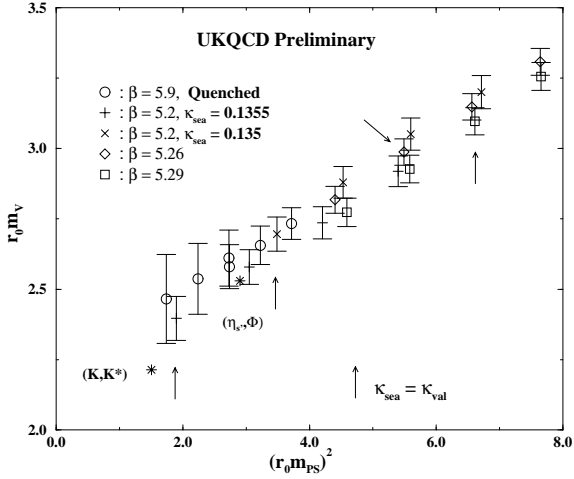}
  \end{center}
\vspace{-1cm}
\caption{Vector mass against pseudoscalar mass squared in units of $r_0$. The arrows indicate those points where $\ksea = \kval$. }
\vspace{-0.75cm}
\label{r0mv}
\end{figure}

\noindent Hadron masses were obtained from correlated least-$\chi^2$ fits. The mesons
have been fitted by a double cosh fit to local and fuzzed correlators
simultaneously. Baryons are fitted by a single exponential fit to
fuzzed correlators only. The ratio of the pseudoscalar to vector
masses is shown in Table~\ref{ratio} for $\ksea = \kval$. One way to
look for dynamical effects in the spectrum is to compare the
pseudoscalar and vector meson masses as $\ksea$ is
varied. Figure~\ref{r0mv} shows a plot of $m_{\rm V}$ against
 $m^2_{\rm PS}$ for all data sets. Since $\beta$ is different for each
data set, the results for the meson masses are shown in units of
$r_0$. Points corresponding to $\ksea = \kval$ are indicated by
arrows. This plot shows that there is a trend towards the experimental
points as $\ksea$ becomes lighter.

Preliminary analysis of the spectrum has been conducted in the
partially quenched scheme where the partially quenched quark mass is
defined as
\be
m_{q}^{\rm PQ} =
\frac{1}{2}\left( \frac{1}{\kval} - \frac{1}{\kappa_{\rm
crit}}\right)
\ee
Here $\kappa_{\rm crit}$ has been determined from an extrapolation
in the improved valence quark mass for each  data set,
$am_{\rm PS}^2 \propto \tilde{m}_{q}(\kval)$, using $b_{\rm m}$ from
perturbation theory.

\begin{figure}
\begin{center}
    \leavevmode
    \includegraphics[angle=0,width=2.8in]{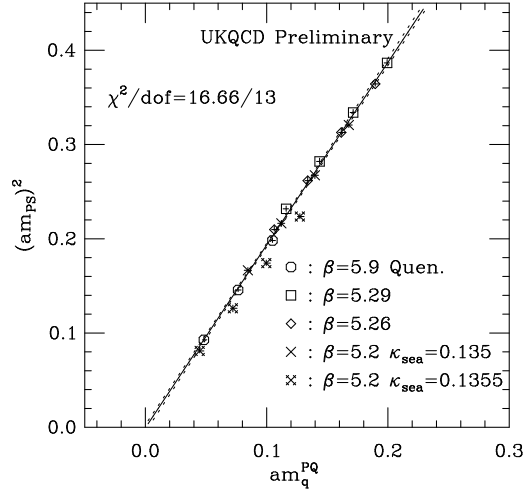}
  \end{center}
\label{pseudo}
\vspace{-1cm}
\caption{Pseudoscalar extrapolation as a function of the partially quenched quark mass in lattice units.}
\vspace{-0.5cm}
\end{figure}

The pseudoscalar extrapolation as a function of $m_{q}^{\rm PQ}$ is
shown for all data sets in Figure~\ref{pseudo}. A straight line has
been fitted to the matched data sets, including the quenched simulation,
using an uncorrelated fit. Data points from the lightest $\ksea$
simulation have been included in the plot. These points clearly have a
different slope from the matched data sets. 
\section{Conclusions}

\noindent We have seen some evidence of screening in the static quark
potential for dynamical ${\rm N_{\rm f}} = 2$ simulations. Using data sets
which have been matched to have the same effective lattice volume
helps to disentangle the screening effects from the discretisation
errors in the potential. Extrapolations of the pseudoscalar mass as a
function of the partially quenched quark mass, show that the slope is
 consistent for the matched ensembles. Further analysis of this
type of extrapolation is in progress.


\end{document}